\documentclass[
]{ceurart}

\usepackage{listings}
\begin{document}

\copyrightyear{2022}
\copyrightclause{Copyright for this paper by its authors.
  Use permitted under Creative Commons License Attribution 4.0
  International (CC BY 4.0).}

\conference{The first annual workshop on Learnersourcing: Student-generated Content @ Scale,
  June 01, 2022, NYC, NY}

\title{Coverage of Course Topics in Learnersourced SQL Exercises}


\author[1]{Nea Pirttinen}[
orcid=0000-0001-5249-5162,
email=nea.pirttinen@helsinki.fi,
]
\address[1]{University of Helsinki, Helsinki, Finland}

\author[2]{Arto Hellas}[
orcid=0000-0001-6502-209X,
email=arto.hellas@aalto.fi,
]
\address[2]{Aalto University, Espoo, Finland}

\author[2]{Juho Leinonen}[
orcid=0000-0001-6829-9449,
email=juho.2.leinonen@aalto.fi,
]


\begin{abstract}
    Learnersourcing is a common task in modern computing classrooms, where it is used, for example, for the creation of educational resources such as multiple-choice questions and programming exercises. One less studied type of learnersourced artefact is SQL exercises. In this work, we explore how well different SQL topics are covered by learnersourced SQL exercises. Covering most course topics would allow students to practice the full content of the course by completing learnersourced exercises. Our results suggest that learnersourcing can be used to create a large pool of SQL exercises that cover most of the topics of the course.
\end{abstract}


\maketitle

\section{Introduction}

Traditionally, learnersourcing has been used, for example, for creating multiple-choice questions (MCQs)~\cite{denny2008peerwise,moore2021examining,singh2021s,khosravi2019ripple}. In computing classrooms, learnersourcing is increasingly used for the creation of programming exercises~\cite{pirttinen2018crowdsourcing,denny2011codewrite}. In our recent prior work, we introduced a system for crowdsourcing SQL exercises~\cite{leinonen2020crowdsourcing}, which to the best of our knowledge is the first such system.

One concrete goal of learnersourcing systems is to create a vast repository of exercises that could complement or even replace instructor-created exercises. An important question related to this is whether student-created exercises cover all the course topics, which would be necessary to allow students to practice all learning objectives of the course. Prior work has studied the coverage of course topics of learnersourced MCQs and found that student-created exercises did cover all course topics~\cite{denny2009coverage}.

In this work-in-progress paper, we report preliminary results related to how well learnersourced SQL exercises cover different SQL concepts. Our goal is to examine whether -- by using mostly learnersourced exercises -- students will get to practice the concepts they are expected to learn during the course.










\section{Tool and Context}

\subsection{SQL Trainer}

SQL Trainer is a learnersourcing system for practicing SQL queries \cite{leinonen2020crowdsourcing}. When using the system, a student is presented with a list of topics in the order of appearance in the course material. Students can choose to either create exercises or practice a specific topic. If the student chooses to create an exercise for a particular topic, the system will ask which database the student wishes to use (selection is made from a pre-defined list of databases created by the course instructor). The student is then required to give a name for the exercise, an exercise description, and a sample solution for that exercise. Once the exercise is completed, it is added to the pool of exercises for that topic. If the student chooses to practice a topic, they will be given a randomly selected exercise that they have not completed yet from the pool of created exercises.

A more detailed description of SQL Trainer and its usage, as well as the description of the teacher view, is presented in \cite{leinonen2020crowdsourcing}.

\subsection{Context and Research Questions}

For the present study, we use data collected from three introductory database courses offered by the University of Helsinki, where SQL trainer was used to support SQL practice. In the courses, SQL trainer had 11 topics (briefly outlined in Table~\ref{tab:topic-ref}). Approximately 10\% of the course grade was based on completing at least four exercises per course topic and creating at least a single exercise per course topic. Students were free to complete and create more exercises than required for the points, however.


Our research questions are as follows:

\begin{itemize}
    \item[\textbf{RQ1.}] Which topics do students create exercises for?
    \item[\textbf{RQ2.}] How well do the exercises created by students cover the course topics?
\end{itemize}

To answer research question 1 and 2, we partially replicate earlier work by Denny et al. \cite{denny2009coverage} and Purchase et al. \cite{purchase2010quality} by studying to what extent the exercises that students complete and create cover the course topics. In particular, we study how many exercises are created by students per course topic, and how the used SQL concepts match the topics.

\section{Results and Discussion}


\begin{table}[hbt!]
\caption{Course topics. Arranged in the order the topics appear in the cource material.}
\label{tab:topic-ref}
\begin{tabular}{l|l}
\textbf{\#} & \textbf{Topic}                                     \\ \hline
1  & Select from a table                       \\ \hline
2  & Filter and order data from a single table \\ \hline
3  & Select from multiple tables               \\ \hline
4  & Select from even more tables              \\ \hline
5  & Other types of joins                      \\ \hline
6  & Adding and removing tables                \\ \hline
7  & Adding data to a database                 \\ \hline
8  & Using update and delete                   \\ \hline
9  & Using functions                           \\ \hline
10 & Using group by and functions              \\ \hline
11 & Functions, group by, having, order        
\end{tabular}
\end{table}

\subsection{Created Exercises}

\begin{table}[]
\caption{The number of exercises created per topic, 11247 in total. Arranged in the order the topics appear in the course material.}
\label{tab:created-exercise-topics}
\begin{tabular}{l|l}
\textbf{Topic}                                           & \textbf{Exercises created} \\ \hline
1                     & 1301 (11.6 \%)                         \\
2 & 1205 (10.7 \%)                         \\
3                     & 1113 (9.9 \%)                         \\
4                & 1048 (9.3 \%)                        \\
5                          & 666 (5.9 \%)                          \\
6                      & 1063 (9.5 \%)                        \\
7                      & 1047 (9.3 \%)                        \\
8                       & 1037 (9.2 \%)                         \\
9                                & 1028 (9.1 \%)                        \\
10                   & 948 (8.4 \%)                         \\
11                       & 879 (7.8 \%)                         \\
\end{tabular}
\end{table}


In total, 1569 students entered the system. Out of these, a total of 1187 created at least one exercise of their own. In total, students created 11247 exercises for 11 different instructor-defined topics. The number of created exercises for each topic can be found in Table \ref{tab:created-exercise-topics} (topic number references listed in Table \ref{tab:topic-ref}).

From Table~\ref{tab:created-exercise-topics}, we can see that there is a clear downwards trend in the number of created exercises per topic. One possible reason for this is that the latter topics come later in the course and thus fewer students actively participate at that point in the course. Another potential explanation for the trend is that the latter topics are related to somewhat more difficult topics and thus students might not be as inclined to create exercises for those. One clear outlier is topic 5, which was related to ``other types of joins''. Why this topic elicited fewer learnersourced exercises requires further research.




\subsection{Topic Coverage}


\begin{figure}[hbt!]
    \centering
    \includegraphics[width=0.5\textwidth]{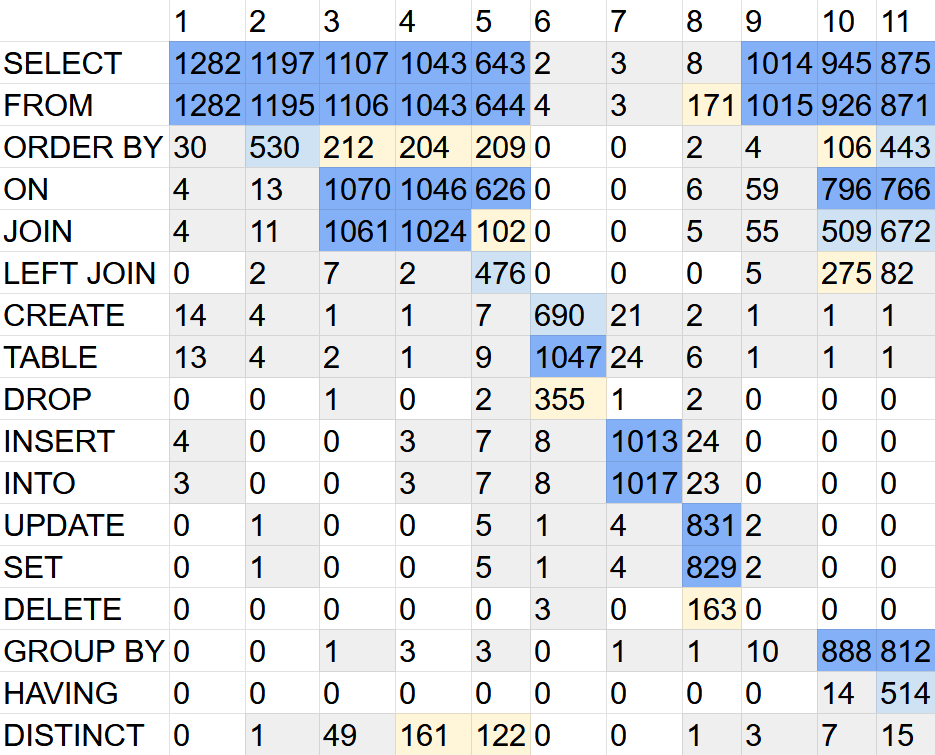}
    \caption{
    The number of SQL concepts used per topic. Color legend: concept used in over 80 \% of the created exercises in dark blue, over 40 \% light blue, over 10 \% yellow, under 10 \% gray, and 0 occurrences white.
    }
    \label{fig:topic-coverage}
\end{figure}

Figure \ref{fig:topic-coverage} presents the most common SQL concepts for each topic (topic number references listed in Table \ref{tab:topic-ref}). Only concepts that were relatively common (occurred in over 10\% of created exercises in at least a single topic) are included. In the vast majority of cases, students are able to identify relevant concepts for the topic, and utilize them when creating an exercise.


Interestingly, based on the SQL concepts used in topic 5, students seem to majorly prefer creating exercises related to left joins compared to right or inner joins (which occurred rarely enough that they were not included in the figure). Similarly, for topic 6, we can see that students are more inclined to create exercises related to creating tables than dropping tables.








\subsection{Future Work}


Altogether, these preliminary results support the use of learnersourcing for creating a large exercise pool with a good coverage of course topics. This provides the opportunity for students to practice a wide variety of SQL concepts using the learnersourced exercises. In our future work, we are interested in extending this research to similarly examine how much students practice the different concepts using the system. In addition, we are keen to study whether the exercises created by students cover varying levels of difficulty and if the exercises are effective. Lastly, we are exploring whether there are differences between demographic groups in to what extent they participate in learnersourcing activities.

\begin{acknowledgments}
We are grateful for the doctoral research grant awarded by Jenny and Antti Wihuri Foundation to the first author.
\end{acknowledgments}

\bibliography{ref}


\end{document}